# Variable selection for the multicategory SVM via adaptive sup-norm regularization

**Hao Helen Zhang**

*Department of Statistics*
*North Carolina State University*
*Raleigh, NC 27695*
*e-mail:* hzhang2@stat.ncsu.edu

**Yufeng Liu**[*]

*Department of Statistics and Operations Research*
*Carolina Center for Genome Sciences*
*University of North Carolina*
*Chapel Hill, NC 27599*
*e-mail:* yfliu@email.unc.edu

**Yichao Wu**

*Department of Operations Research and Financial Engineering*
*Princeton University*
*Princeton, NJ 08544*
*e-mail:* yichaowu@princeton.edu

**Ji Zhu**

*Department of Statistics*
*University of Michigan*
*Ann Arbor, MI 48109*
*e-mail:* jizhu@umich.edu

**Abstract:** The Support Vector Machine (SVM) is a popular classification paradigm in machine learning and has achieved great success in real applications. However, the standard SVM can not select variables automatically and therefore its solution typically utilizes all the input variables without discrimination. This makes it difficult to identify important predictor variables, which is often one of the primary goals in data analysis. In this paper, we propose two novel types of regularization in the context of the multicategory SVM (MSVM) for simultaneous classification and variable selection. The MSVM generally requires estimation of multiple discriminating functions and applies the argmax rule for prediction. For each individual variable, we propose to characterize its importance by the supnorm of its coefficient vector associated with different functions, and then minimize the MSVM hinge loss function subject to a penalty on the sum of supnorms. To further improve the supnorm penalty, we propose the adaptive regularization, which allows different weights imposed on different variables according to their relative importance. Both types of regularization automate variable selection in the process of building classifiers, and lead to sparse multi-classifiers with enhanced interpretability and improved accuracy, especially for high dimensional low sample size data. One big advantage of the supnorm penalty is its easy implementation via standard linear programming. Several simulated examples and one real gene data analysis demonstrate the outstanding performance of the adaptive supnorm penalty in various data settings.

[*]Corresponding author. The authors thank the editor Professor Larry Wasserman, the associate editor, and two reviewers for their constructive comments and suggestions. Liu's research was supported in part by the National Science Foundation DMS-0606577 and DMS-0747575. Wu's research was supported by the National Institute of Health NIH R01-GM07261. Zhang's research was supported in part by the National Science Foundation DMS-0645293 and the National Institute of Health NIH/NCI R01 CA-085848. Zhu's research was supported in part by the National Science Foundation DMS-0505432 and DMS-0705532.

## Contents

## 1. Introduction

In supervised learning problems, we are given a training set of $n$ examples from $K \geq 2$ different populations. For each example in the training set, we observe its covariate $\mathbf{x}_i \in \mathbb{R}^d$ and the corresponding label $y_i$ indicating its membership. Our ultimate goal is to learn a classification rule which can accurately predict the class label of a future example based on its covariate. Among many classification methods, the Support Vector Machine (SVM) has gained much popularity in both machine learning and statistics. The seminal work by Vapnik (1995, 1998) has laid the foundation for the general statistical learning theory and the SVM, which furthermore inspired various extensions on the SVM. For other references on the binary SVM, see Christianini and Shawe-Taylor (2000), Schölkopf and Smola (2002), and references therein. Recently a few attempts have been made to generalize the SVM to multiclass problems, such as Vapnik (1998), Weston and Watkins (1999), Crammer and Singer (2001), Lee et al. (2004), Liu and Shen (2006), and Wu and Liu (2007a).

While the SVM outperforms many other methods in terms of classification accuracy in numerous real problems, the implicit nature of its solution makes it less attractive in providing insight into the predictive ability of individual variables. Often times, selecting relevant variables is the primary goal of data mining. For the binary SVM, Bradley and Mangasarian (1998) demonstrated the utility of the $L_1$ penalty, which can effectively select variables by shrinking small or redundant coefficients to zero. Zhu et al. (2003) provides an efficient algorithm to compute the entire solution path for the $L_1$-norm SVM. Other forms of penalty have also been studied in the context of binary SVMs, such as the $L_0$ penalty (Weston et al., 2003), the SCAD penalty (Zhang et al., 2006), the $L_q$ penalty (Liu et al., 2007), the combination of $L_0$ and $L_1$ penalty (Liu and Wu, 2007), the combination of $L_1$ and $L_2$ penalty (Wang et al., 2006), the $F_\infty$ norm (Zou and Yuan, 2006), and others (Zhao et al., 2006; Zou, 2006).

For multiclass problems, variable selection becomes more complex than the binary case, since the MSVM requires estimation of multiple discriminating functions, among which each function has its own subset of important predictors. One natural idea is to extend the $L_1$ SVM to the $L_1$ MSVM, as done in the recent work of Lee et al. (2006) and Wang and Shen (2007b). However, the $L_1$ penalty does not distinguish the source of coefficients. It treats all the coefficients equally, no matter whether they correspond to the same variable or different variables, or they are more likely to be relevant or irrelevant. In this paper, we propose a new regularized MSVM for more effective variable selection. In contrast to the $L_1$ MSVM, which imposes a penalty on the sum of absolute values of all coefficients, we penalize the sup-norm of the coefficients associated with each variable. The proposed method is shown to be able to achieve a higher degree of model parsimony than the $L_1$ MSVM without compromising classification accuracy.

This paper is organized as follows. Section 2 formulates the sup-norm regularization for the MSVM. Section 3 proposes an efficient algorithm to implement the MSVM. Section 4 discusses an adaptive approach to improve performance of the sup-norm MSVM by allowing different penalties for different covariates according to their relative importance. Numerical results on simulated and gene expression data are given in Sections 5 and 6, followed by a summary.

## 2. Methodology

In $K$-category classification problems, we code $y$ as $\{1, \ldots, K\}$ and define $\mathbf{f} = (f_1, \ldots, f_K)$ as a decision function vector. Each $f_k$, a mapping from the input domain $\mathbb{R}^d$ to $\mathbb{R}$, represents the strength of the evidence that an example with input $\mathbf{x}$ belongs to the class $k$; $k = 1, \ldots, K$. A classifier induced by $\mathbf{f}$,

$$\phi(\mathbf{x}) = \arg \max_{k=1,\ldots,K} f_k(\mathbf{x}),$$

assigns an example with $\mathbf{x}$ to the class with the largest $f_k(\mathbf{x})$. We assume the $n$ training pairs $\{(\mathbf{x}_i, y_i), i = 1, \ldots, n\}$ are independently and identically distributed according to an unknown probability distribution $P(\mathbf{x}, y)$. Given a classifier $\mathbf{f}$, its performance is measured by the generalization error, $\text{GE}(\mathbf{f}) = P(Y \neq \arg\max_k f_k(\mathbf{X})) = E_{(\mathbf{X},Y)}[I(Y \neq \arg\max_k f_k(\mathbf{X}))]$.

Let $p_k(\mathbf{x}) = \Pr(Y = k|\mathbf{X} = \mathbf{x})$ be the conditional probability of class $k$ given $\mathbf{X} = \mathbf{x}$. The Bayes rule which minimizes the GE is then given by

$$\phi_B(\mathbf{x}) = \arg\min_{k=1,\ldots,K}[1 - p_k(\mathbf{x})] = \arg\max_{k=1,\ldots,K} p_k(\mathbf{x}). \quad (2.1)$$

For nonlinear problems, we assume $f_k(\mathbf{x}) = b_k + \sum_{j=1}^{q} w_{kj} h_j(\mathbf{x})$ using a set of basis functions $\{h_j(\mathbf{x})\}$. This linear representation of a nonlinear classifier through basis functions will greatly facilitate the formulation of the proposed method. Alternatively nonlinear classifiers can also be achieved by applying the kernel trick (Boser et al., 1992). However, the kernel classifier is often given as a black box function, where the contribution of each individual covariate to the decision rule is too implicit to be characterized. Therefore we will use the basis expansion to construct nonlinear classifiers in the paper.

The standard multicategory SVM (MSVM; Lee et al., 2004) solves

$$\min_{\mathbf{f}} \frac{1}{n} \sum_{i=1}^{n} \sum_{k=1}^{K} I(y_i \neq k)[f_k(\mathbf{x}_i) + 1]_+ + \lambda \sum_{k=1}^{K} \sum_{j=1}^{d} w_{kj}^2, \quad (2.2)$$

under the sum-to-zero constraint $\sum_{k=1}^{K} f_k = 0$. The sum-to-zero constraint used here is to follow Lee et al. (2004) in their framework for the MSVM. It is imposed to eliminate redundancy in $f_k$'s and to assure identifiability of the solution. This constraint is also a necessary condition for the Fisher consistency of the MSVM proposed by Lee et al. (2004). To achieve variable selection, Wang and Shen (2007b) proposed to impose the $L_1$ penalty on the coefficients and the corresponding $L_1$ MSVM then solves

$$\min_{\mathbf{b},\mathbf{w}} \frac{1}{n} \sum_{i=1}^{n} \sum_{k=1}^{K} I(y_i \neq k)[b_k + \mathbf{w}_k^T \mathbf{x}_i + 1]_+ + \lambda \sum_{k=1}^{K} \sum_{j=1}^{d} |w_{kj}| \quad (2.3)$$

under the sum-to-zero constraint. For linear classification rules, we start with $f_k(\mathbf{x}) = b_k + \sum_{j=1}^{d} w_{kj} x_j$, $k = 1, \ldots, K$. The sum-to-zero constraint then becomes

$$\sum_{k=1}^{K} b_k = 0, \quad \sum_{k=1}^{K} w_{kj} = 0, \quad j = 1, \ldots, d. \quad (2.4)$$

The $L_1$ MSVM treats all $w_{kj}$'s equally without distinction. As opposed to this, we take into account the fact that some of the coefficients are associated with the same covariate, therefore it is more natural to treat them as a group rather than separately.

Define the weight matrix $W$ of size $K \times d$ such that its $(k, j)$ entry is $w_{kj}$. The structure of $W$ is shown as follows:

|         | $x_1$    | $\cdots$ | $x_j$    | $\cdots$ | $x_d$    |
|---------|----------|----------|----------|----------|----------|
| Class 1 | $w_{11}$ | $\cdots$ | $w_{1j}$ | $\cdots$ | $w_{1d}$ |
|         | $\cdots$ | $\cdots$ | $\cdots$ | $\cdots$ | $\cdots$ |
| Class k | $w_{k1}$ | $\cdots$ | $w_{kj}$ | $\cdots$ | $w_{kd}$ |
|         | $\cdots$ | $\cdots$ | $\cdots$ | $\cdots$ | $\cdots$ |
| Class K | $w_{K1}$ | $\cdots$ | $w_{Kj}$ | $\cdots$ | $w_{Kd}$ |

Throughout the paper, we use $\mathbf{w}_k = (w_{k1},\ldots,w_{kd})^{\mathrm{T}}$ to represent the $k$th row vector of $W$, and $\mathbf{w}_{(j)} = (w_{1j},\ldots,w_{Kj})^{\mathrm{T}}$ for the $j$th column vector of $W$. According to Crammer and Singer (2001), the value $b_k + \mathbf{w}_k^{\mathrm{T}}\mathbf{x}$ defines the similarity score of the class $k$, and the predicted label is the index of the row attaining the highest similarity score with $\mathbf{x}$. We define the sup-norm for the coefficient vector $\mathbf{w}_{(j)}$ as

$$\|\mathbf{w}_{(j)}\|_\infty = \max_{k=1,\cdots,K} |w_{kj}|. \tag{2.5}$$

In this way, the importance of each covariate $x_j$ is directly controlled by its largest absolute coefficient. We propose the sup-norm regularization for MSVM:

$$\min_{\mathbf{b},\mathbf{w}} \quad \frac{1}{n}\sum_{i=1}^n \sum_{k=1}^K I(y_i \neq k)[b_k + \mathbf{w}_k^T \mathbf{x}_i + 1]_+ + \lambda \sum_{j=1}^d \|\mathbf{w}_{(j)}\|_\infty,$$

$$\text{subject to} \quad \mathbf{1}^T\mathbf{b} = 0, \quad \mathbf{1}^T\mathbf{w}_{(j)} = 0, \quad \text{for } j=1,\ldots,d, \tag{2.6}$$

where $\mathbf{b} = (b_1,\ldots,b_K)^{\mathrm{T}}$.

The sup-norm MSVM encourages more sparse solutions than the $L_1$ MSVM, and identifies important variables more precisely. In the following, we describe the main motivation of the sup-norm MSVM, which makes it more attractive for variable selection than the $L_1$ MSVM. Firstly, with a sup-norm penalty, a noise variable is removed if and only if all corresponding $K$ estimated coefficients are 0. On the other hand, if a variable is important with a positive sup-norm, the sup-norm penalty, unlike the $L_1$ penalty, does not put any additional penalties on the other $K-1$ coefficients. This is desirable since a variable will be kept in the model as long as the sup-norm of the $K$ coefficients is positive. No further shrinkage is needed for the remaining coefficients in terms of variable selection. For illustration, we plot the region $0 \leq t_1 + t_2 \leq C$ in Figure 1, where $t_1 = \max(w_{11},w_{21},w_{31},w_{41})$ and $t_2 = \max(w_{12},w_{22},w_{32},w_{42})$. Clearly, the sup-norm penalty shrinks sum of two maximums corresponding to two variables. This helps to lead to more parsimonious models. In short, in contrast to the $L_1$ penalty, the sup-norm utilizes the group information of the decision function vector and consequently the sup-norm MSVM can deliver better variable selection.

For three-class problems, we show that the $L_1$ MSVM and the new proposed sup-norm MSVM give identical solutions after adjusting the tuning parameters, which is due to the sum-to-zero constraints on $\mathbf{w}_{(j)}$'s. This equivalence, however, does not hold for the adaptive procedures introduced in Section 4.

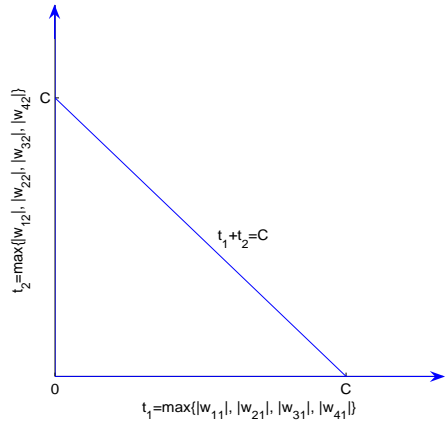

Fig 1. *Illustrative plot of the shrinkage property of the sup-norm.*

PROPOSITION 2.1. *When $K = 3$, the $L_1$ MSVM (2.3) and the sup-norm MSVM (2.6) are equivalent.*

When $K > 3$, our empirical experience shows that the sup-norm MSVM generally performs well in terms of classification accuracy.

Here we would like to point out two fundamental differences between the sup-norm penalty and the $F_\infty$ penalty used for group variable selection (Zhao et al., 2006; Zou and Yuan, 2006) considering their similar expressions. The purpose of group selection is to select several prediction variables altogether if these predictors work as a group. Therefore, each $F_\infty$ term in Zou and Yuan (2006) is based on the regression coefficients of several variables which belong to one group, whereas each supnorm penalty in (2.6) is associated with only one prediction variable. Secondly, in the implementation of the $F_\infty$, one has to decide in advance the number of groups and which variables belong to a certain group, whereas in the supnorm SVM each variable is naturally associated with its own group and the number of groups is same as the number of covariates.

As a remark, we point out that Argyriou et al. (2007, 2006) proposed a similar penalty for the purpose of multi-task feature learning. Specifically, they used a mixture of $L_1$ and $L_2$ penalties. They first applied the $L_2$ penalty for each feature across different tasks and then used the $L_1$ penalty for feature selection. In contrast, our penalty is a combination of the $L_1$ and supnorm penalties for multicategory classification.

The tuning parameter $\lambda$ in (2.6) balances the tradeoff between the data fit and the model parsimony. A proper choice of $\lambda$ is important to assure good performance of the resulting classifier. If the chosen $\lambda$ is too small, the procedure tends to overfit the training data and gives a less sparse solution; on the other hand, if $\lambda$ is too large, the solution can become very sparse but possibly with a low prediction power. The choice of the tuning parameter is typically done by minimizing either an estimate of generalization error or other related performance measures. In simulations, we generate an extra independent tuning set to choose the best $\lambda$. For real data analysis, we use leave-one-out cross validation of the misclassification rate to select $\lambda$.

## 3. Computational Algorithms

In this section we show that the optimization problem (2.6) can be converted to a linear programming (LP) problem, and can therefore be solved using standard LP techniques in polynomial time. This great computational advantage is very important in real applications, especially for large data sets.

Let $A$ be an $n \times K$ matching matrix with its entry $a_{ik} = I(y_i \neq k)$ for $i = 1, \ldots, n$ and $k = 1, \ldots, K$. First we introduce slack variables $\xi_{ik}$ such that

$$\xi_{ik} = \left[b_k + \mathbf{w}_k^T \mathbf{x}_i + 1\right]_+ \quad \text{for} \quad i = 1, \ldots, n; \quad k = 1, \ldots, K. \quad (3.1)$$

The optimization problem (2.6) can be expressed as

$$\min_{\mathbf{b}, \mathbf{w}, \boldsymbol{\xi}} \quad \frac{1}{n} \sum_{i=1}^n \sum_{k=1}^K a_{ik}\xi_{ik} + \lambda \sum_{j=1}^d \|\mathbf{w}_{(j)}\|_\infty,$$

$$\text{subject to} \quad \mathbf{1}^T \mathbf{b} = 0, \quad \mathbf{1}^T \mathbf{w}_{(j)} = 0, \quad j = 1, \ldots, d,$$

$$\xi_{ik} \geq b_k + \mathbf{w}_k^T \mathbf{x}_i + 1, \quad \xi_{ik} \geq 0, \quad i = 1, \ldots, n; \quad k = 1, \ldots, K. \quad (3.2)$$

To further simplify (3.2), we introduce a second set of slack variables

$$\eta_j = \|\mathbf{w}_{(j)}\|_\infty = \max_{k=1,\ldots,K} |w_{kj}|,$$

which add some new constraints to the problem:

$$|w_{kj}| \leq \eta_j, \quad \text{for} \quad k = 1, \ldots, K; \quad j = 1, \ldots, d.$$

Finally write $w_{kj} = w_{kj}^+ - w_{kj}^-$, where $w_{kj}^+$ and $w_{kj}^-$ denote the positive and negative parts of $w_{kj}$, respectively. Similarly, $\mathbf{w}_j^+$ and $\mathbf{w}_j^-$ respectively consist of the positive and negative parts of components in $\mathbf{w}_j$. Denote $\boldsymbol{\eta} = (\eta_1, \ldots, \eta_d)^T$; then (3.2) becomes

$$\min_{\mathbf{b}, \mathbf{w}, \boldsymbol{\xi}, \boldsymbol{\eta}} \quad \frac{1}{n} \sum_{i=1}^n \sum_{k=1}^K a_{ik}\xi_{ik} + \lambda \sum_{j=1}^d \eta_j,$$

$$\text{subject to} \quad \mathbf{1}^T \mathbf{b} = 0, \quad \mathbf{1}^T [\mathbf{w}_{(j)}^+ - \mathbf{w}_{(j)}^-] = 0, \quad j = 1, \ldots, d,$$

$$\xi_{ik} \geq b_k + [\mathbf{w}_k^+ - \mathbf{w}_k^-]^T \mathbf{x}_i + 1, \quad \xi_{ik} \geq 0, \quad i = 1, \ldots, n; \quad k = 1, \ldots, K,$$

$$\mathbf{w}_{(j)}^+ + \mathbf{w}_{(j)}^- \leq \boldsymbol{\eta}, \quad \mathbf{w}_{(j)}^+ \geq \mathbf{0}, \quad \mathbf{w}_{(j)}^- \geq \mathbf{0}, \quad j = 1, \ldots, d. \quad (3.3)$$

## 4. Adaptive Penalty

In (2.3) and (2.6), the same weights are used for different variables in the penalty terms, which may be too restrictive, since a smaller penalty may be more desired for those variables which are so important that we want to retain them in the model. In this section, we suggest that different variables should be penalized differently according to their relative importance. Ideally, large penalties should be imposed on redundant variables in order to eliminate them from models more easily; and small penalties should be used on important variables in order to retain them in the final classifier. Motivated by this, we consider the following adaptive $L_1$ MSVM:

$$\min_{\mathbf{b},\mathbf{w}} \quad \frac{1}{n}\sum_{i=1}^{n}\sum_{k=1}^{K} I(y_i \neq k)[b_k + \mathbf{w}_k^T \mathbf{x}_i + 1]_+ + \lambda \sum_{k=1}^{K}\sum_{j=1}^{d} \tau_{kj}|w_{kj}|,$$

$$\text{subject to} \quad \mathbf{1}^T \mathbf{b} = 0, \quad \mathbf{1}^T \mathbf{w}_{(j)} = 0, \quad \text{for } j=1,\ldots,d, \qquad (4.1)$$

where $\tau_{kj} > 0$ represents the weight for coefficient $w_{kj}$.

Adaptive shrinkage for each variable has been proposed and studied in various contexts of regression problems, including the adaptive LASSO for linear regression (Zou, 2006), proportional hazard models (Zhang and Lu, 2007), and quantile regression (Wang et al., 2007; Wu and Liu, 2007b). In particular, Zou (2006) has established the oracle property of the adaptive LASSO and justified the use of different amounts of shrinkage for different variables. Due to the special form of the sup-norm SVM, we consider the following two ways to employ the adaptive penalties:

[I]

$$\min_{\mathbf{b},\mathbf{w}} \quad \frac{1}{n}\sum_{i=1}^{n}\sum_{k=1}^{K} I(y_i \neq k)[b_k + \mathbf{w}_k^T \mathbf{x}_i + 1]_+ + \lambda \sum_{j=1}^{d} \tau_j \|\mathbf{w}_{(j)}\|_\infty,$$

$$\text{subject to} \quad \mathbf{1}^T \mathbf{b} = 0, \quad \mathbf{1}^T \mathbf{w}_{(j)} = 0, \quad \text{for } j=1,\ldots,d, \qquad (4.2)$$

[II]

$$\min_{\mathbf{b},\mathbf{w}} \quad \frac{1}{n}\sum_{i=1}^{n}\sum_{k=1}^{K} I(y_i \neq k)[b_k + \mathbf{w}_k^T \mathbf{x}_i + 1]_+ + \lambda \sum_{j=1}^{d} \|(\boldsymbol{\tau}\mathbf{w})_{(j)}\|_\infty,$$

$$\text{subject to} \quad \mathbf{1}^T \mathbf{b} = 0, \quad \mathbf{1}^T \mathbf{w}_{(j)} = 0, \quad \text{for } j=1,\ldots,d, \qquad (4.3)$$

where the vector $(\boldsymbol{\tau}\mathbf{w})_{(j)} = (\tau_{1j}w_{1j},\ldots,\tau_{Kj}w_{Kj})^\mathrm{T}$ for $j=1,\ldots,d$.

In (4.1), (4.2), and (4.3), the weights can be regarded as leverage factors, which are adaptively chosen such that large penalties are imposed on coefficients of unimportant covariates and small penalties on coefficients of important ones. Let $\tilde{\mathbf{w}}$ be the solution to standard MSVM (2.2) with the $L_2$ penalty. Our empirical experience suggests that

$$\tau_{kj} = \frac{1}{|\tilde{w}_{kj}|}$$

is a good choice for (4.1) and (4.3), and

$$\tau_j = \frac{1}{\|\tilde{\mathbf{w}}_{(j)}\|_\infty}$$

is a good choice for (4.2). If $\tilde{w}_{kj} = 0$, which implies the infinite penalty on $w_{kj}$, we set the corresponding coefficient solution $\hat{w}_{kj}$ to be zero.

In terms of computational issues, all three problems (4.1), (4.2), and (4.3) can be solved as LP problems. Their entire solution paths may be obtained by some modifications of the algorithms in Wang and Shen (2007b).

## 5. Simulation

In this section, we demonstrate the performance of six MSVM methods: the standard $L_2$ MSVM, $L_1$ MSVM, sup-norm MSVM, adaptive $L_1$ MSVM, and the two adaptive sup-norm MSVMs. Three simulation models are considered: (1) a linear example with five classes; (2) a linear example with four classes; (3) a nonlinear example with three classes. In each simulation setting, $n$ observations are simulated as the training data, and another $n$ observations are generated for tuning the regularization parameter $\lambda$ for each procedure. Therefore the total sample size is $2n$ for obtaining the final classifiers. To test the accuracy of the classification rules, we also independently generate $n'$ observations as a test set. The tuning parameter $\lambda$ is selected via a grid search over the grid: $\log_2(\lambda) = -14, -13, \ldots, 15$. When a tie occurs, we choose the larger value of $\lambda$. As we suggest in Section 4, we use the $L_2$ MSVM solution to derive the weights in the adaptive MSVMs. The $L_2$ MSVM solution is the final tuned solution using the separate tuning set. Once the weights are chosen, we tune the parameter $\lambda$ in the adaptive procedure via the tuning set.

We conduct 100 simulations for each classification method under all settings. Each fitted classifier is then evaluated in terms of its classification accuracy and variable selection performance. For each method, we report its average testing error, the number of correct and incorrect zero coefficients among $Kd$ coefficients, the model size as the number of important ones among the $d$ variables, and the number of times that the true model is correctly identified. The numbers given in the parentheses in the tables are the standard errors of the testing errors. We also summarize the frequency of each variable being selected over 100 runs. All simulations are done using the optimization software CPLEX with the AMPL interface (Fourer et al., 2003). More information about CPLEX can be found on the ILOG website http://www.ilog.com/products/optimization/.

### 5.1. Five-Class Example

Consider a five-class example, with the input vector **x** in a 10-dimensional space. The first two components of the input vector are generated from a mixture

Gaussian in the following way: for each class $k$, generate $(x_1, x_2)$ independently from $N(\boldsymbol{\mu}_k, \sigma_1^2 I_2)$, with

$$\boldsymbol{\mu}_i = 2\left(\cos([2k-1]\pi/5), \sin([2k-1]\pi/5)\right), \quad k = 1, 2, 3, 4, 5,$$

and the remaining eight components are i.i.d. generated from $N(0, \sigma_2^2)$. We generate the same number of observations in each class. Here $\sigma_1 = \sqrt{2}, \sigma_2 = 1, n = 250$, and $n' = 50,000$.

TABLE 1
*Classification and variable selection results for the five-class example. TE, CZ, IZ, MS, and CM refer to the testing error, the number of correct zeros, the number of incorrect zeros, the model size, and the number of times that the true model is correctly identified, respectively.*

| Method | TE | CZ | IZ | MS | CM |
|---|---|---|---|---|---|
| L2 | 0.454 (0.034) | 0.00 | 0.00 | 10.00 | 0 |
| L1 | 0.558 (0.022) | 24.88 | 2.81 | 6.60 | 21 |
| Adapt-L1 | 0.553 (0.020) | 30.23 | 2.84 | 5.14 | 40 |
| Supnorm | 0.453 (0.020) | 33.90 | 0.01 | 3.39 | 68 |
| Adapt-supI | 0.455 (0.024) | 39.92 | 0.01 | 2.08 | 98 |
| Adapt-supII | 0.457 (0.046) | 39.40 | 0.09 | 2.17 | 97 |
| Bayes | 0.387 (—) | 41 | 0 | 2 | 100 |

Table 1 shows that, in terms of classification accuracy, the $L_2$ MSVM, the supnorm MSVM, and the two adaptive supnorm MSVMs are among the best and their testing errors are close to each other. In terms of other measurements such as the number of correct/incorrect zeros, the model size, and the number of times that the true model is correctly identified, the supnorm MSVM procedures work much better than other MSVM methods.

Table 2 shows the frequency of each variable being selected by each procedure in 100 runs. The type I sup-norm MSVM performs the best among all. Overall the adaptive MSVMs show significant improvement over the non-adaptive classifiers in terms of both classification accuracy and variable selection.

TABLE 2
*Variable selection frequency results for the five-class example.*

| | Selection Frequency | | | | | | | | | |
|---|---|---|---|---|---|---|---|---|---|---|
| Method | $x_1$ | $x_2$ | $x_3$ | $x_4$ | $x_5$ | $x_6$ | $x_7$ | $x_8$ | $x_9$ | $x_{10}$ |
| L2 | 100 | 100 | 100 | 100 | 100 | 100 | 100 | 100 | 100 | 100 |
| L1 | 100 | 100 | 59 | 55 | 60 | 58 | 56 | 61 | 57 | 54 |
| Adapt-L1 | 100 | 100 | 44 | 40 | 43 | 37 | 39 | 41 | 35 | 35 |
| Supnorm | 100 | 100 | 15 | 17 | 20 | 17 | 14 | 20 | 17 | 19 |
| Adapt-supI | 100 | 100 | 1 | 1 | 0 | 2 | 1 | 1 | 1 | 1 |
| Adapt-supII | 100 | 100 | 2 | 2 | 2 | 2 | 2 | 2 | 3 | 2 |

### 5.2. Four-Class Linear Example

In the simulation example in Section 5.1, the informative variables are important for all classes. In this section, we consider an example where the informative vari-ables are important for some classes but not important for other classes. Specifically, we generate four i.i.d important variables $x_1, x_2, x_3, x_4$ from Unif$[-1, 1]$ as well as six independent i.i.d noise variables $x_5, \ldots, x_{10}$ from $N(0, 8^2)$. Define the functions

$$\begin{aligned} f_1 &= -5x_1 + 5x_4, \\ f_2 &= 5x_1 + 5x_2, \\ f_3 &= -5x_2 + 5x_3, \\ f_4 &= -5x_3 - 5x_4, \end{aligned}$$

and set $p_k(\mathbf{x}) = P(Y = k | X = \mathbf{x}) \propto \exp(f_k(\mathbf{x})), k = 1, 2, 3, 4$. In this example, we set $n = 200$ and $n' = 40,000$. Note that $x_1$ is not important for distinguishing class 3 and class 4. Similarly, $x_2$ is noninformative for class 1 and class 4, $x_3$ is noninformative for class 1 and class 2, and $x_4$ is noninformative for class 2 and class 3.

Table 3
*Classification and variable selection results for the four-class linear example.*

| Method | TE | CZ | IZ | MS | CM |
|---|---|---|---|---|---|
| L2 | 0.336 (0.063) | 0.0000 | 0.0000 | 10.00 | 0 |
| L1 | 0.340 (0.069) | 2.5100 | 0.1600 | 9.99 | 0 |
| Adapt-L1 | 0.320 (0.079) | 18.2300 | 0.2600 | 7.21 | 21 |
| Supnorm | 0.332 (0.070) | 0.8500 | 0.1400 | 9.98 | 0 |
| Adapt-supI | 0.327 (0.076) | 9.3300 | 0.1400 | 7.83 | 15 |
| Adapt-supII | 0.326 (0.071) | 9.9000 | 0.1400 | 7.69 | 9 |
| Bayes | 0.1366 (—) | 32 | 0 | 4 | 100 |

Table 3 summarizes the performance of various procedures, and Table 4 shows the frequency of each variable being selected by each procedure in 100 runs. Due to the increased difficulty of this problem, the performances of all methods are not as good as that of the five-class example. From these results, we can see that the adaptive procedures work better than the non-adaptive procedures both in terms of both classification accuracy and variable selection. Furthermore, the adaptive $L_1$ MSVM performs the best overall. This is due to the difference between the $L_1$ and the supnorm penalties. Our proposed supnorm penalty treats all coefficients of one variable corresponding to different classes as a group and removes the variable if it is non-informative across all class labels. By design of this example, important variables have zero coefficients for certain classes. As a result, our supnorm penalty does not deliver the best performance. Nevertheless, the adaptive supnorm procedures still perform reasonably.

Table 4
*Variable selection frequency results for the four-class example.*

| Method | $x_1$ | $x_2$ | $x_3$ | $x_4$ | $x_5$ | $x_6$ | $x_7$ | $x_8$ | $x_9$ | $x_{10}$ |
|---|---|---|---|---|---|---|---|---|---|---|
| | | | | | Selection Frequency | | | | | |
| L2 | 100 | 100 | 100 | 100 | 100 | 100 | 100 | 100 | 100 | 100 |
| L1 | 100 | 100 | 100 | 100 | 100 | 99 | 100 | 100 | 100 | 100 |
| Adapt-L1 | 100 | 100 | 100 | 100 | 55 | 53 | 59 | 56 | 49 | 49 |
| Supnorm | 100 | 100 | 100 | 100 | 100 | 99 | 100 | 100 | 100 | 99 |
| Adapt-supI | 100 | 100 | 100 | 100 | 67 | 64 | 71 | 60 | 58 | 63 |
| Adapt-supII | 100 | 100 | 100 | 100 | 65 | 66 | 65 | 58 | 56 | 59 |

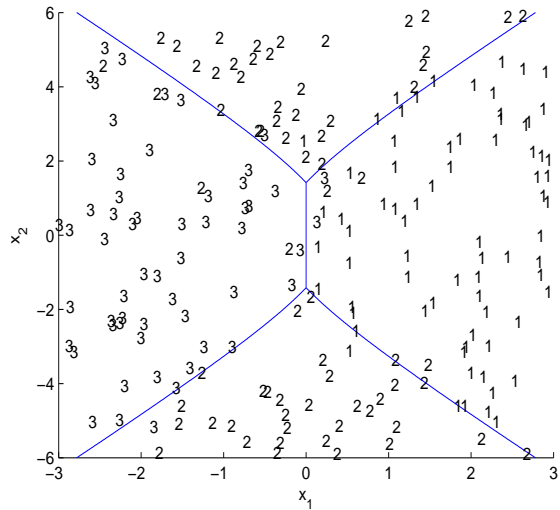

Fig 2. *The Bayes boundary for the nonlinear three-class example.*

### 5.3. Nonlinear Example

In this nonlinear 3-class example, we first generate $x_1 \sim \text{Unif}[-3, 3]$ and $x_2 \sim \text{Unif}[-6, 6]$. Define the functions

$$\begin{aligned} f_1 &= -2x_1 + 0.2x_1^2 - 0.1x_2^2 + 0.2, \\ f_2 &= -0.4x_1^2 + 0.2x_2^2 - 0.4, \\ f_3 &= 2x_1 + 0.2x_1^2 - 0.1x_2^2 + 0.2, \end{aligned}$$

and set $p_k(\mathbf{x}) = P(Y = k|X = \mathbf{x}) \propto \exp(f_k(\mathbf{x})), k = 1, 2, 3$. The Bayes boundary is plotted in Figure 2. We also generate three noise variables $x_i \sim N(0, \sigma^2)$, $i = 3, 4, 5$. In this example, we set $\sigma = 2$ and $n' = 40,000$.

To achieve nonlinear classification, we fit the nonlinear MSVM by including the five main effects, their square terms, and their cross products as the basis functions. The results with $n = 200$ are summarized in Tables 5 and 6. Clearly, the adaptive $L_1$ SVM and the two adaptive sup-norm SVMs deliver more accurate and sparse classifiers than the other methods. In this example, there are correlations among covariates and consequently the variable selection task becomes more challenging. This difficulty is reflected in the variable selection frequency reported in Table 6. Despite the difficulty, the adaptive procedures are able to remove noise variables reasonably well.

TABLE 5
*Classification and variable selection results using second order polynomial basis functions for the nonlinear example in Section 5.3 with $n = 200$.*

| Method | TE | CZ | IZ | MS | CM |
|---|---|---|---|---|---|
| L2 | 0.167 (0.013) | 0.00 | 0.00 | 20.00 | 0 |
| L1 | 0.151 (0.012) | 21.42 | 0.03 | 14.91 | 0 |
| Adapt-L1 | 0.140 (0.010) | 43.13 | 0.00 | 6.92 | 31 |
| Supnorm | 0.150 (0.012) | 22.70 | 0.01 | 14.43 | 0 |
| Adapt-supI | 0.140 (0.010) | 40.84 | 0.00 | 7.21 | 31 |
| Adapt-supII | 0.140 (0.011) | 41.50 | 0.00 | 6.21 | 36 |
| Bayes | 0.120 (—) | 52 | 0 | 3 | 100 |

TABLE 6
*Variable selection frequency results for the nonlinear example using second order polynomial basis functions with $n = 200$.*

| | Selection Frequency | | | | | | | | | |
|---|---|---|---|---|---|---|---|---|---|---|
| Method | $x_1$ | $x_1^2$ | $x_2^2$ | $x_2$ | $x_3$ | $x_4$ | $x_5$ | $x_3^2$ | $x_4^2$ | $x_5^2$ |
| L2 | 100 | 100 | 100 | 100 | 100 | 100 | 100 | 100 | 100 | 100 |
| L1 | 100 | 100 | 100 | 69 | 44 | 50 | 43 | 80 | 84 | 89 |
| Adapt-L1 | 100 | 100 | 100 | 33 | 21 | 21 | 20 | 24 | 18 | 22 |
| Supnorm | 100 | 100 | 100 | 67 | 37 | 42 | 34 | 84 | 80 | 75 |
| Adapt-supI | 100 | 100 | 100 | 31 | 21 | 21 | 26 | 21 | 25 | 24 |
| Adapt-supII | 100 | 100 | 100 | 22 | 18 | 12 | 19 | 18 | 16 | 18 |
| | $x_1x_2$ | $x_1x_3$ | $x_1x_4$ | $x_1x_5$ | $x_2x_3$ | $x_2x_4$ | $x_2x_5$ | $x_3x_4$ | $x_3x_5$ | $x_4x_5$ |
| L2 | 100 | 100 | 100 | 100 | 100 | 100 | 100 | 100 | 100 | 100 |
| L1 | 80 | 55 | 57 | 65 | 86 | 88 | 90 | 69 | 72 | 70 |
| Adapt-L1 | 31 | 20 | 18 | 20 | 28 | 26 | 31 | 20 | 17 | 22 |
| Supnorm | 79 | 62 | 58 | 55 | 87 | 89 | 91 | 62 | 68 | 73 |
| Adapt-supI | 31 | 22 | 17 | 28 | 30 | 29 | 30 | 24 | 16 | 25 |
| Adapt-supII | 25 | 15 | 14 | 19 | 30 | 23 | 22 | 16 | 17 | 17 |

To examine the performance of various methods using a richer set of basis functions, we also fit nonlinear MSVMs via polynomial basis of degree 3 with 55 basis functions. Results of classification and variable selection with $n = 200$ and 400 are reported in Tables 7 and 8 respectively. Compared with the case of the second order polynomial basis, classification testing errors using the third order polynomial basis are much larger for the $L_2$, $L_1$, and supnorm MSVMs, but similar for the adaptive procedures. Due to the large basis set, none of the methods can identify the correct model. However, the adaptive procedures can eliminate more noise variables than the non-adaptive procedures. This further demonstrates the effectiveness of adaptive weighting. The results of variable selection frequency (not reported due to lack of space) show a similar pattern as that of the second order polynomial. When $n$ increases from 200 and 400, we can see that classification accuracy for all methods increases as expected. Interestingly, compared to the case of $n = 200$, the performance of variable selection with $n = 400$ for non-adaptive procedures stays relatively the same, while improves dramatically for the adaptive procedures.

TABLE 7
Classification and variable selection results using third order polynomial basis functions for the nonlinear example in Section 5.3 with $n = 200$.

| Method | TE | CZ | IZ | MS | CM |
|---|---|---|---|---|---|
| L2 | 0.213 (0.018) | 0.00 | 0.00 | 55.00 | 0 |
| L1 | 0.170 (0.015) | 59.22 | 0.57 | 40.44 | 0 |
| Adapt-L1 | 0.138 (0.015) | 120.71 | 0.17 | 19.28 | 0 |
| Supnorm | 0.171 (0.015) | 60.08 | 0.61 | 40.06 | 0 |
| Adapt-supI | 0.141 (0.016) | 114.29 | 0.17 | 20.22 | 0 |
| Adapt-supII | 0.142 (0.015) | 106.78 | 0.22 | 19.75 | 0 |
| Bayes | 0.120 (—) | 157 | 0 | 3 | 100 |

TABLE 8
Classification and variable selection results using third order polynomial basis functions for the nonlinear example in Section 5.3 with $n = 400$.

| Method | TE | CZ | IZ | MS | CM |
|---|---|---|---|---|---|
| L2 | 0.162 (0.008) | 0.00 | 0.00 | 55.00 | 0 |
| L1 | 0.143 (0.008) | 60.13 | 0.34 | 40.50 | 0 |
| Adapt-L1 | 0.124 (0.004) | 139.71 | 0.00 | 11.01 | 0 |
| Supnorm | 0.144 (0.010) | 60.51 | 0.32 | 40.24 | 0 |
| Adapt-supI | 0.125 (0.005) | 139.41 | 0.00 | 10.37 | 0 |
| Adapt-supII | 0.125 (0.004) | 132.96 | 0.00 | 10.96 | 0 |
| Bayes | 0.120 (—) | 157 | 0 | 3 | 100 |

## 6. Real Example

DNA microarray technology has made it possible to monitor mRNA expressions of thousands of genes simultaneously. In this section, we apply our six different MSVMs on the children cancer data set in Khan et al. (2001). Khan et al. (2001) classified the small round blue cell tumors (SRBCTs) of childhood into 4 classes; namely neuroblastoma (NB), rhabdomyosarcoma (RMS), non-Hodgkin lymphoma (NHL), and the Ewing family of tumors (EWS) using cDNA gene expression profiles. After filtering, 2308 gene profiles out of 6567 genes are given in the data set, available at http://research.nhgri.nih.gov/microarray/Supplement/. The data set includes a training set of size 63 and a test set of size 20. The distributions of the four distinct tumor categories in the training and test sets are given in Table 9. Note that Burkitt lymphoma (BL) is a subset of NHL.

To analyze the data, we first standardize the data sets by applying a simple linear transformation based on the training data. Specifically, we standardize the expression $\tilde{x}_{gi}$ of the $g$-th gene of subject $i$ to obtain $x_{gi}$ by the following formula:

$$x_{gi} = \frac{\tilde{x}_{gi} - \frac{1}{n}\sum_{j=1}^{n} \tilde{x}_{gj}}{sd(\tilde{x}_{g1}, \cdots, \tilde{x}_{gn})}.$$

TABLE 9
*Class distribution of the microarray example.*

| Data set | NB | RMS | BL | EWS | Total |
|---|---|---|---|---|---|
| Training | 12 | 20 | 8 | 23 | 63 |
| Test | 6 | 5 | 3 | 6 | 20 |

TABLE 10
*Classification results of the microarray data using 200 genes.*

| | | Selected genes | |
|---|---|---|---|
| Penalty | Testing Error | Top 100 | Bottom 100 |
| L2 | 0 | 100 | 100 |
| L1 | 1/20 | 62 | 1 |
| Adp-L1 | 0 | 53 | 1 |
| Supnorm | 1/20 | 53 | 0 |
| Adp-supI | 1/20 | 50 | 0 |
| Adp-supII | 1/20 | 47 | 0 |

Then we rank all genes using their marginal relevance in class separation by adopting a simple criterion used in Dudoit et al. (2002). Specifically, the relevance measure for gene $g$ is defined to be the ratio of between classes sum of squares to within class sum of squares as follows:

$$R(g) = \frac{\sum_{i=1}^{n} \sum_{k=1}^{K} I(y_i = k)(\bar{x}_{\cdot g}^{(k)} - \bar{x}_{\cdot g})^2}{\sum_{i=1}^{n} \sum_{k=1}^{K} I(y_i = k)(x_{ig} - \bar{x}_{\cdot g}^{(k)})^2}, \quad (6.1)$$

where $n$ is the size of the training set, $\bar{x}_{\cdot g}^{(k)}$ denotes the average expression level of gene $g$ for class $k$ observations, and $\bar{x}_{\cdot g}$ is the overall mean expression level of gene $g$ in the training set. To examine the performance of variable selection of all different methods, we select the top 100 and bottom 100 genes as covariates according the relevance measure $R$. Our main goal here is to get a set of "important" genes and also a set of "unimportant" genes, and to see whether our methods can effectively remove the "unimportant" genes.

All six MSVMs with different penalties are applied to the training set. We use leave-one-out cross validation on the standardized training data with 200 genes for the purpose of tuning parameter selection and then apply the resulting classifiers on the testing data. The results are tabulated in Table 10. All methods have either 0 or 1 misclassification on the testing set. In terms of gene selection, three sup-norm MSVMs are able to eliminate all bottom 100 genes and they use around 50 genes out of the top 100 genes to achieve comparable classification performance to other methods.

In Figure 3, we plot heat maps of both training and testing sets on the left and right panels respectively. In these heat maps, rows represent 50 genes selected by the Type I sup-norm MSVM and columns represent patients. The gene expression values are reflected by colors on the plot, with red representing the highest expression level and blue the lowest expression level. For visualization, we group columns within each class together and use hierarchical clustering with correlation distance on the training set to order the genes so that genes close to each other have similar expressions. From the left panel on Figure 3, we can observe four block structures associated with four classes. This implies that the 50 genes selected are highly informative in predicting the tumor types. For the testing set shown on the right panel, we can still see the four blocks although the structure and pattern are not as clean as the training set. It is interesting to note that several genes in the testing set have higher expression levels, i.e., more red, than the training set. In summary, we conclude that the proposed sup-norm MSVMs are indeed effective in performing simultaneous classification and variable selection.

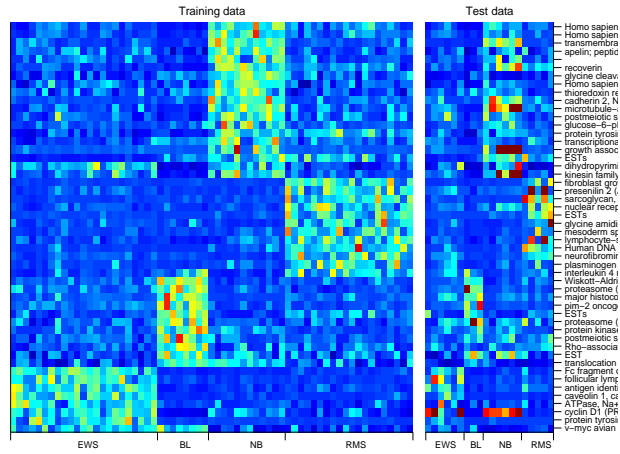

FIG 3. *Heat maps of the microarray data. The left and right panels represent the training and testing sets respectively.*

## 7. Discussion

As pointed out in Lafferty and Wasserman (2006), sparse learning is an important but challenging issue for high dimensional data. In this paper, we propose a new regularization method which applies the sup-norm penalty to the MSVM to achieve variable selection. Through the new penalty, the natural group effect of the coefficients associated with the same variable is embedded in the regularization framework. As a result, the sup-norm MSVMs can perform better variable selection and deliver more parsimonious classifiers than the $L_1$ MSVMs. Moreover, our results show that the adaptive procedures work very well and improve the corresponding nonadaptive procedures. The adaptive $L_1$ procedure can in some settings be as good as and sometimes better than the adaptive supnorm procedures. As a future research direction, we will further investigate the theoretical properties of proposed methods.

In some problems, it is possible to form groups among covariates. As argued in Yuan and Lin (2006) and Zou and Yuan (2006), it is advisable to use such group information in the model building process to improve accuracy of the prediction. The notion of "group lasso" has also been studied in the context of learning a kernel (Micchelli and Pontil, 2007). If such kind of group information is available for multicategory classification, there will be two kinds of group information available for model building, one type of group formed by the same covariate corresponding to different classes as considered in the paper and the other kind formed among covariates. A future research direction is to combine both group information to construct a new multicategory classification method. We believe that such potential classifiers can outperform those without using the additional information.

This paper focuses on the variable selection issue for supervised learning. In practice, semi-supervised learning is often encountered, and many methods have been developed including Zhu et al. (2003) and Wang and Shen (2007a). Another future topic is to generalize the sup-norm penalty to the context of semi-supervised learning.

## Appendix

**Proof of Proposition 2.1:** Without loss of generality, assume that $\{w_{1j}, w_{2j}, w_{3j}\}$ are all nonzero. Because of the sum-to-zero constraint $w_{1j} + w_{2j} + w_{3j} = 0$, there must be one component out of $\{w_{1j}, w_{2j}, w_{3j}\}$ has a different sign from the other two. Suppose the sign of $w_{1j}$ differs from the other two and then $|w_{1j}| = |w_{2j}| + |w_{3j}|$ by the sum-to-zero constraint. Consequently, we have $|w_{1j}| = \max\{|w_{1j}|, |w_{2j}|, |w_{3j}|\}$. Therefore, $\sum_{k=1}^{3} |w_{kj}| = 2\|\mathbf{w}_{(j)}\|_\infty$. The equivalence of problem (2.2) with the tuning parameter $\lambda$ and problem (2.6) with the tuning parameter $2\lambda$ can be then established. This completes the proof.